\documentclass[letterpaper,onecolumn,11pt]{article} 
\usepackage[margin=1in]{geometry}
\usepackage{palatino}
\usepackage[usenames]{color}
\usepackage{graphicx}
\usepackage{epsfig}
\usepackage{fancyvrb}
\usepackage{url}
\usepackage{cite}
\usepackage{multirow}
\usepackage{wrapfig}
\usepackage{xspace}
\usepackage{comment}

\newcommand{\cnt}[1]{\multicolumn{1}{|c|}{#1}}

\begin{document}

\title{\bf \LARGE{ARBAC Policy for a Large Multi-National Bank}}

\author{Karthick Jayaraman\\Syracuse University\\kjayaram@syr.edu \and
  Vijay Ganesh\\MIT\\vganesh@csail.mit.edu \and Mahesh
  Tripunitara\\University of Waterloo\\tripunit@uwaterloo.ca
 \and Martin C. Rinard\\MIT\\mrinard@csail.mit.edu 
  \and Steve J. Chapin\\Syracuse University\\chapin@syr.edu}

\date{\today}
\maketitle

\begin{abstract}
  Administrative role-based access control (ARBAC) is the first
  comprehensive administrative model proposed for role-based access
  control (RBAC). ARBAC has several features for designing highly
  expressive policies, but current work has not highlighted the
  utility of these expressive policies. In this report, we present a
  case study of designing an ARBAC policy for a bank comprising 18
  branches. Using this case study we provide an assessment about the
  features of ARBAC that are likely to be used in realistic policies.
\end{abstract}

\section{Introduction}

This case study describes the design of an administrative role-based
access-control (ARBAC) policy for a bank comprising 18 branches. We
designed the ARBAC policy to meet the following two goals:
\begin{enumerate}
\item Facilitate the business functions in each branch by
  appropriately provisioning roles to facilitate the tasks of each job
  position.

\item Enforce a separation of privilege (SOP) property to prevent
  collusive behavior in the execution of important job tasks. 
\end{enumerate}

The business functions and job roles used in each bank branch are
based on an existing study.  Schadd et al.~\cite{bank} describes a
role-based access-control (RBAC) policy for an European bank
branch. We extend this policy into an ARBAC policy in two
steps. First, we add 17 additional branches with the same functions
and job roles. This is because we limited our case study to 18
branches, which is approximately 600 roles based on the number of
roles in each branch.  Second, we designed \emph{can\_assign} and
\emph{can\_revoke} rules for administering role assignment actions
such that the SOP property is enforced.

SOP is a key concern for financial institutions because they are
required to enforce such properties either by regulators or to be
compliant with standards such as ISO 9000~\cite{cprh}. SOP has its
primary objective in the prevention of fraud or collusive behavior in
crucial operations. Typically, SOP is enforced by dividing tasks and
privileges for executing an operation among multiple users, and making
sure that a single user cannot obtain all the privileges necessary for
independently completing an operation.

For example, let us consider that the \emph{Widget} corporation wants
to enforce a high level of transparency in the processing of purchase
orders. To meet this objective, \emph{Widget} splits the purchase
order transaction into two tasks, namely creation and approval. The
permissions for the tasks are assigned such that junior level
employees have the ability to create purchase orders, but only the
division's manager can approve the purchase orders. Formally, the SOP
constraint for this example can be stated as, $\langle$ \{ Creation,
Approval\}, 1 $\rangle$, which means that a user can at most
have one role from the set \{ Creation, Approval\}.

This case study illustrates how SOP constraints can be enforced by
using a well designed ARBAC policy. Three features of ARBAC, namely
\emph{disjunctions}, \emph{positive preconditions}, and \emph{mixed
  roles}, are useful in expressing the SOP constraints. This case
study illustrates how these features are used in the policy. These
features are complexity sources with respect to analyzing these
policies for safety. Therefore, use of these features makes the
automatic analysis of ARBAC policies harder. However, because of the
utility of these features in enforcing properties such as SOP, we
envision that realistic policies will take advantage of these
features.

This case study also describes how to formulate safety queries for
verifying properties of the policy. Formally, a safety query is a
tuple of the form $\langle$ \emph{u, r} $\rangle$ that questions
whether a user \emph{u} can be assigned to a role \emph{r}. Several
questions about the policy can be formulated as one or more safety
queries~\cite{scott07,jha07}. We illustrate this with examples of
safety queries for verifying the SOP property. Model checkers could be
used for verifying these safety questions prior to deploying the
policy.

The remainder of this case study is structured as follows. Section
\ref{sec:roles} describes the roles and the role hierarchy in the
policy. Section \ref{sec:rules} describes the design of the
\emph{can\_assign} and \emph{can\_revoke} rules. Section
\ref{sec:questions} describes how to formulate analysis
questions. Section \ref{sec:conclusion} provides a summary.

\section{Roles and Role Hierarchy}
\label{sec:roles}

\begin{figure}
\centering
\includegraphics[scale=0.4]{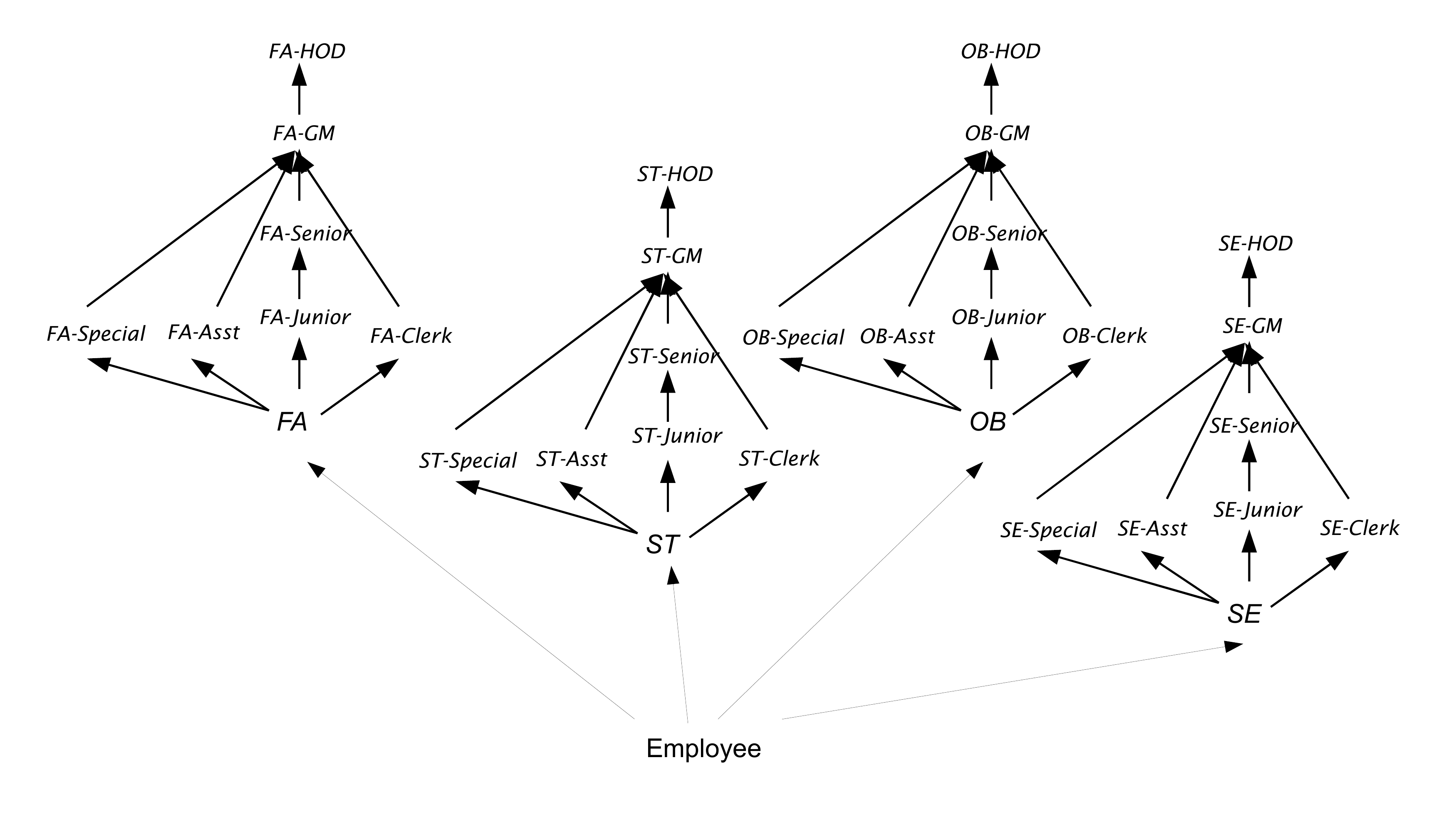}
\caption{Role Hierarchy Design}
\label{fig:hierarchy}
\end{figure}

The bank comprises 18 branches. Each branch in the bank has 33 roles
that are spread over four business divisions, namely financial analyst
(FA), share technician (ST), office banking (OB), and support
e-commerce (SE). Table \ref{tab:roles} contains the list of roles in
each branch and Figure \ref{fig:hierarchy} contains the role
hierarchy. There are eight roles per business division, comprising the
following:
\begin{enumerate}
\item A role for each business division from which all the other roles
  in the business division inherit. For example, all the roles in the
  FA business division inherit from the FA role.
  \item Two managerial roles. For example, FA-HOD and FA-GM are
    managerial roles in the FA division.
  \item Five non-managerial roles. For example, FA-Asst,
    FA-Specialist, FA-Senior, FA-Junior, and FA-Clerk are
    non-managerial positions in the FA division.
\end{enumerate}
Each branch has a role called employee, from which all other
branch-specific roles inherit. Each branch has the same set of roles,
leading to 594 roles in the bank policy comprising 18 branches. 

\begin{table}
\centering
\begin{tabular}{|l|l|l|l|}
\hline
 & \cnt{Function} & \cnt{Position} & \cnt{Role} \\
\hline
1. & Financial Analyst & Head of Division & FA-HOD \\
2. & Financial Analyst & Group Manager & FA-GM \\
3. & Financial Analyst & Specialist & FA-Special \\
4. & Financial Analyst & Assistant & FA-Asst \\
5. & Financial Analyst & Senior & FA-Senior \\
6. & Financial Analyst & Junior & FA-Junior \\
7. & Financial Analyst & Clerk & FA-Clerk \\
8. & Share Technician & Head of Division & ST-HOD \\
9. & Share Technician & Group Manager & ST-GM \\
10. & Share Technician & Specialist & ST-Special \\
11. & Share Technician & Assistant & ST-Asst \\
12. & Share Technician & Senior & ST-Senior \\
13. & Share Technician & Junior & ST-Junior \\
14. & Share Technician & Clerk & ST-Clerk \\
15. & Office Banking & Head of Division & OB-HOD \\
16. & Office Banking & Group Manager & OB-GM \\
17. & Office Banking & Specialist & OB-Special \\
18. & Office Banking & Assistant & OB-Asst \\
19. & Office Banking & Senior	& OB-Sr \\
20. & Office Banking & Junior & OB-Jr \\
21. & Office Banking & Clerk & OB-Clerk \\
22. & Support E-Comm & Head of Division & SE-HOD \\
23. & Support E-Comm & Group Manager & SE-GM \\
24. & Support E-Comm & Specialist & SE-Special \\
25. & Support E-Comm & Assistant & SE-Asst \\
26. & Support E-Comm & Senior & SE-Sr \\
27. & Support E-Comm & Junior & SE-Jr \\
28. & Support E-Comm & Clerk & SE-Clerk\\
29. & Financial Analyst & - & FA \\
30. & Share Technician & - & ST \\
31. & Office Banking & - & OB \\
32. & Support E-Comm & - & SE \\
33. & Branch Employee & - & Employee \\
\hline
\end{tabular}
\caption{Roles Derived from Function and Official Positions}
\label{tab:roles}
\end{table}

In each branch, the five non-managerial roles in each business
division have a separation of privilege constraint such that a user
may not be assigned to more than 3 roles of the five roles. For
example, FA-Asst, FA-Specialist, FA-Senior, FA-Junior, and FA Clerk
are the five junior roles in the FA division. A user who is already
assigned to one of these roles can be additionally assigned to at most
2 roles out of the remainder four roles.

Our policy assumes separate administration, which implies that
administrative roles are not managed by the same set of rules that
apply to the regular roles. We have a single role named \emph{Admin}
used for administering role assignments and revocations. Therefore,
the \emph{Admin} role appears as a precondition in all the assignment
and revocation rules.

\section{Can\_Assign and Can\_Revoke Rules}
\label{sec:rules}

We designed \emph{can\_assign} rules to enforce the constraint that a
user can be assigned to at most 3 roles out of the five non-managerial
roles in each business division.  This constraint can be stated
formally as,

\begin{center} $\langle$ \{FA-Asst, FA-Specialist, FA-Senior, FA-Junior, FA
Clerk\}, 3 $\rangle$ \end{center} 

The \emph{can\_assign} rules are designed to express the valid
conditions for assigning a user to a particular role. The conditions
specify the role memberships that are required for user to be assigned
to a role. To meet our SOP objective, we need to enumerate all the
valid conditions that can entitle a user to be assigned to each of the
five non-managerial role. These conditions can then be expressed as
one or more \emph{can\_assign} rules.

We illustrate the design of the \emph{can\_assign} rules using an
example. For example, let us consider the FA-Clerk role. Table
\ref{tab:ca} contains the \emph{can\_assign rules} for the role
FA-Clerk. A user can be assigned to the role FA-Clerk, under three
cases:

\begin{enumerate}
\item \emph{Case 1:} If a user is is not assigned to any other
  managerial role, then he can be assigned to the FA-Clerk role. To
  express this condition, we created a \emph{can\_assign} rule that
  contains FA as a positive precondition and four negative
  preconditions for the other four non-managerial roles.

\item \emph{Case 2:} If a user is already assigned to a single
  non-managerial role, then the user may be assigned to an additional
  non-managerial role. To express this condition, we need four
  \emph{can\_assign} rules. Each of these four rules will have 2
  positive preconditions and 3 negative preconditions. Of the two
  positive preconditions, one is for a business division role FA and
  the other is one of the four non-managerial roles. The remainder
  non-managerial roles appear as negative preconditions.

\item \emph{Case 3:} If a user is already assigned to two
  non-managerial roles, then the user can be assigned to an additional
  non-managerial role. To express this condition, we need six
  \emph{can\_assign} rules. Each of the six rules have 3 positive
  preconditions and 2 negative preconditions. The 3 positive
  preconditions include the business division role and two of the four
  non-managerial roles. The remainder non-managerial roles appear as
  negative preconditions.
\end{enumerate}

As illustrated by Table \ref{tab:ca}, the \emph{can\_assign} rules
make use of disjunctions, positive preconditions, and mixed roles. The
valid conditions for assigning the FA-Clerk role is essentially a
disjunction, in which each \emph{can\_assign} rule is a
disjunct. Also, several of the \emph{can\_assign} rules have mixed
preconditions because they have both positive and negative
preconditions.

We followed the same procedure for designing the \emph{can\_assign}
rules for all the other non-managerial roles.  The complete list of
the \emph{can\_assign} and \emph{can\_revoke} rules can be obtained
from our policy
file\footnote{\url{http://kjayaram.mysite.syr.edu/mohawk/Mohawk.html}}.

\begin{table}
\centering
{\footnotesize
\textsf{
\begin{tabular}{|r|c|c|c|}
  \hline
  & \bf{Admin Role} & \bf{Preconditions} & \bf{Target Role} \\
  \hline
  1. & Admin & FA$\wedge \neg$FA-Asst$\wedge
  \neg$FA-Specialist$\wedge \neg$FA-Senior$\wedge \neg$FA-Junior & FA-Clerk \\
  2.& Admin & FA$\wedge$FA-Asst$\wedge
  \neg$FA-Specialist$\wedge \neg$FA-Senior$\wedge \neg$FA-Junior & FA-Clerk \\
  3. & Admin & FA$\wedge \neg$FA-Asst$\wedge$FA-Specialist$\wedge \neg$FA-Senior$\wedge \neg$FA-Junior & FA-Clerk \\
  4. & Admin & FA$\wedge \neg$FA-Asst$\wedge
  \neg$FA-Specialist$\wedge$FA-Senior$\wedge \neg$FA-Junior & FA-Clerk \\
  5. & Admin & FA$\wedge \neg$FA-Asst$\wedge
  \neg$FA-Specialist$\wedge \neg$FA-Senior$\wedge$FA-Junior & FA-Clerk \\
  6. &  Admin & FA$\wedge \neg$FA-Asst$\wedge
  \neg$FA-Specialist$\wedge$FA-Senior$\wedge$FA-Junior & FA-Clerk \\
  7. & Admin & FA$\wedge \neg$FA-Asst$\wedge$FA-Specialist$\wedge \neg$FA-Senior$\wedge$FA-Junior & FA-Clerk \\
  8. & Admin & FA$\wedge
  \neg$FA-Asst$\wedge$FA-Specialist$\wedge$FA-Senior$\wedge \neg$FA-Junior & FA-Clerk \\
  9. & Admin & FA$\wedge$FA-Asst$\wedge
  \neg$FA-Specialist$\wedge \neg$FA-Senior$\wedge$FA-Junior & FA-Clerk \\
  10. & Admin & FA$\wedge$FA-Asst$\wedge
  \neg$FA-Specialist$\wedge$FA-Senior$\wedge \neg$FA-Junior & FA-Clerk \\
  11. & Admin & FA$\wedge$FA-Asst$\wedge$FA-Specialist$\wedge\neg$FA-Senior$\wedge\neg$FA-Junior & FA-Clerk \\
  \hline
\end{tabular}
}
}
\caption{Can\_Assign rules for FA-Clerk Roles}
\label{tab:ca}
\end{table}

The can\_assign rules for the managerial roles were designed to
enforce that a user assigned to any of the non-managerial roles cannot
be assigned to a managerial role. Therefore, assignment rules for the
managerial roles had negative preconditions for all the non-managerial
roles and one positive precondition for the business division role.

In our policy, all the roles are revocable. Therefore, we had 594
can\_revoke rules, one for each role. We did not see any reason to
make a branch-specific role irrevocable.

\section{Analysis Questions}
\label{sec:questions}

As mentioned earlier, several questions about the policy can be
expressed as a safety query. We illustrate how the following two
analysis questions can be expressed as a safety query:
\begin{enumerate}
\item Can a user be assigned to four non-managerial roles in a
  business division in any of the 18 branches?
\item Can a user be assigned to four non-managerial roles in a
  business division in all the 18 branches?
\end{enumerate}
Both these questions can be encoded as a safety question of the form
$\langle u, targetrole \rangle$ as follows.

To express these analysis questions as safety queries, we need to add
some additional roles, \emph{can\_assign}, and \emph{can\_revoke}
roles. These additions do not affect the valid administrative actions for
the other roles described in the policy. 

For each branch, we add two additional roles, i.e., we add roles
\emph{AnyFour}$_i$ and \emph{Branch}$_i$ for each branch. The
objective of the \emph{AnyFour}$_i$ is to identify if a user can be
assigned to four non-managerial roles. We add \emph{can\_assign} rules
for this role in each branch to express the condition that if a user
is a member of four non-managerial roles, then he may be assigned to
this special role. The \emph{Branch}$_i$ roles help in the encoding of
both safety questions, so we will refer to them as helper roles. The
\emph{can\_assign} rules for each of the helper roles express the
condition that if a user can be assigned to either \emph{AnyFour}$_i$
or \emph{Branch}$_{(i+1)}$, then he may be assigned to
\emph{Branch}$_i$.

To express question (1) as a safety question, we add a single
\emph{can\_assign} rule for $targetrole$ that is of the form $\langle
Admin, branch_1, targetrole \rangle$. The consequence of this rule is
that If a user can be assigned to $targetrole$, then it implies that a
user can be assigned to four non-managerial roles in at least one of
the 18 branches.

\sloppypar{ To express question (2) as a safety question, we add a
  single \emph{can\_assign} rule for $targetrole$ that is of the form
  $\langle Admin, branch_1 \wedge .. \wedge branc_{18}, targetrole
  \rangle$. The consequence of this rule is that if a use can be
  assigned to $targetrole$, then it implies that a user can be
  assigned to four non-managerial roles in all the branches.



\section{Summary}
\label{sec:conclusion}

We illustrated the design of an ARBAC policy with the intent of
enforcing separation of privilege for a bank comprising 18
branches. The separation of privilege constraint that we have used is
emblematic of realistic concerns. Disjunctions, positive
preconditions, and mixed roles are very useful for encoding SOP
constraints. The SOP constraints can be verified by designing safety
queries.

\bibliographystyle{abbrv}
\bibliography{casestudy}

\end{document}